\documentclass[%
 reprint,
superscriptaddress,
 amsmath,amssymb,
 aps,
]{revtex4-2}

\usepackage{graphicx}
\usepackage{dcolumn}
\usepackage{bm}

\usepackage[mathlines]{lineno}
\usepackage{color}

\begin{document}

\preprint{APS/123-QED}

\title{Measurement of the transverse asymmetry of $\gamma$-rays in the $^{117}$Sn(n,$\gamma$)$^{118}$Sn reaction}
 
\author{S. Endo}
\email{endo.shunsuke@jaea.go.jp}
\affiliation{Japan Atomic Energy Agency, 2-4 Shirakata, Tokai, Ibaraki 319-1195, Japan}
\affiliation{Nagoya University, Furocho, Chikusa, Nagoya 464-8062, Japan}
\author{T. Okudaira}
\affiliation{Nagoya University, Furocho, Chikusa, Nagoya 464-8062, Japan}
\affiliation{Japan Atomic Energy Agency, 2-4 Shirakata, Tokai, Ibaraki 319-1195, Japan}
\author{R. Abe}
\affiliation{Nagoya University, Furocho, Chikusa, Nagoya 464-8062, Japan}
\author{H. Fujioka}
\affiliation{Tokyo Institute of Techonology, Meguro, Tokyo 152-8511, Japan}
\author{K. Hirota}\altaffiliation[Present address:]{High Energy Accelerator Research Organization, 1-1 Oho, Tsukuba, Ibaraki 305-0801, Japan.}
\affiliation{Nagoya University, Furocho, Chikusa, Nagoya 464-8062, Japan}
\author{A. Kimura}
\affiliation{Japan Atomic Energy Agency, 2-4 Shirakata, Tokai, Ibaraki 319-1195, Japan}
\author{M. Kitaguchi}
\affiliation{Nagoya University, Furocho, Chikusa, Nagoya 464-8062, Japan}
\author{T. Oku}
\affiliation{Japan Atomic Energy Agency, 2-4 Shirakata, Tokai, Ibaraki 319-1195, Japan}
\affiliation{Ibaraki University, Mito, Ibaraki 310-8512, Japan}
\author{K. Sakai}
\affiliation{Japan Atomic Energy Agency, 2-4 Shirakata, Tokai, Ibaraki 319-1195, Japan}
\author{T. Shima}
\affiliation{Osaka University, Ibaraki, Osaka 567-0047, Japan}
\author{H. M. Shimizu}
\affiliation{Nagoya University, Furocho, Chikusa, Nagoya 464-8062, Japan}
\author{S. Takada}\altaffiliation[Present address:]{Tohoku University, 41 Kawauchi, Aoba-ku, Sendai, 980-8576 Japan.}
\affiliation{Kyushu University, 744 Motooka, Nishi-ku, Fukuoka 819-0395, Japan}
\affiliation{Japan Atomic Energy Agency, 2-4 Shirakata, Tokai, Ibaraki 319-1195, Japan}
\author{S. Takahashi}
\affiliation{Ibaraki University, Mito, Ibaraki 310-8512, Japan}
\author{T. Yamamoto}
\affiliation{Nagoya University, Furocho, Chikusa, Nagoya 464-8062, Japan}
\author{H. Yoshikawa}
\affiliation{Osaka University, Ibaraki, Osaka 567-0047, Japan}
\author{T. Yoshioka}
\affiliation{Kyushu University, 744 Motooka, Nishi-ku, Fukuoka 819-0395, Japan}

\date{\today}

\begin{abstract}
Largely enhanced parity-violating effects observed in compound resonances induced by epithermal neutrons are currently attributed to the mixing of parity-unfavored partial amplitudes in the entrance channel of the compound states. Furthermore, it is proposed that the same mechanism that enhances the parity-violation also enhances the breaking of time-reversal-invariance in the compound nucleus. The entrance-channel mixing induces energy-dependent spin-angular correlations of individual $\gamma$-rays emitted from the compound nuclear state. For a detailed study of the mixing model, a $\gamma$-ray yield in the reaction of $^{117}$Sn(n,$\gamma$)$^{118}$Sn was measured using the pulsed beam of polarized epithermal neutrons and Ge detectors. An angular dependence of asymmetric $\gamma$-ray yields for the orientation of the neutron polarization was observed.
\end{abstract}

\maketitle

\section{Introduction}
In several p-wave resonances of neutron-induced compound nuclear states with the targets of medium-heavy nuclei such as $^{139}$La, $^{117}$Sn, $^{131}$Xe, and others, an extremely large parity-violation is observed, compared with the small parity-violation due to the weak interaction in nucleon-nucleon scattering \cite{Michel}. The enhancement is currently explained as the result of the mixing between s- and p-wave amplitudes in the entrance channel of the compound nuclear states called the s-p mixing model \cite{Flambaum}. It is suggested that the breaking of the time-reversal-invariance (T-violation) is also enhanced in the compound nucleus, same as the enhancement of parity-violation \cite{Gudkov}. A detailed study of the enhancement mechanism is necessary to quantify the applicability in the order-of-magnitude enhancement of T-violation in nucleon-nucleon interactions beyond the standard model of elementary particles.

The enhanced T-violating effects are expected to be accessible in the neutron spin behavior during transmission through a spin-polarized nuclear target. The nuclei listed above are candidates for the nuclear targets in T-violation search experiments because they show an extremely large enhancement of parity-violation in the neutron energy range around eV, expecting the T-violation enhancement is comparable with the parity-violating enhancement.
Among them, $^{117}$Sn is a good candidate because of spin of 1/2, which is advantageous for the nuclear polarization required for the T-violation search.

The differential cross-section of (n,$\gamma$) reactions can be written in the form of an expression by Legendre polynomials:
\begin{eqnarray}
\label{eq:cor1}
    \frac{d\sigma_{\textrm{n}\gamma}}{d\Omega}&&=\frac{1}{2}\left\{a_0+a_1\bm{k}_\textrm{n}\cdot\bm{k}_\gamma+a_2\bm{\sigma}_\textrm{n}\cdot(\bm{k}_\textrm{n}\times\bm{k}_\gamma)\right. \nonumber\\
    &&\left.+a_3\left(\bm{k}_\textrm{n}\cdot\bm{k}_\gamma-\frac{1}{3}\right)\right\},
\end{eqnarray}
where $\bm{k}_\textrm{n}$, $\bm{k}_\gamma$, and $\bm{\sigma}_\textrm{n}$ are unit vectors parallel to the incident neutron momentum, the emitted $\gamma$-rays momentum, and the incident neutron spin, respectively. Higher-order expansion terms are ignored. The measurements of the coefficients of each correlation term, $a_i$, lead to determining the enhancement of the T-violation and to studying the s-p mixing model.

The correlation terms of (n,$\gamma$) reactions have been measured for several nuclei. The neutron energy-dependent angular distribution of $\gamma$-rays, which relates to $a_1$ term, was measured at the 0.74-eV resonance of $^{139}$La by using an intense pulsed neutron beam and a germanium (Ge) detector assembly at beamline 04 of the Materials and Life Science Experimental Facility (MLF) in the J-PARC \cite{Okudaira,Okudaira2021}. Furthermore, the energy-dependent $\gamma$-ray asymmetry of the 0.74-eV resonance with respect to the transverse polarization of incident neutrons, transverse asymmetry, has been measured by installing a neutron polarization device using polarized $^3$He, known as $^3$He spin filter, on the beamline 04 \cite{Yamamoto}. This transverse asymmetry is expressed by $a_2$ term.

In the case of $^{117}$Sn, the angular distribution, $a_1$ term, of the 1.3-eV resonance has been measured using the same experimental and analysis method of $^{139}$La \cite{Koga}. 
Skoy et al. measured the transverse asymmetry of the 1.3-eV resonance for only one $\gamma$-ray emitted angle with a NaI detector in the IBR-30 reactor and a neutron polarization device using a polarized proton target, which tends to generate neutron backgrounds scattered with proton nuclei \cite{Skoi}. They defined and analyzed the asymmetry, which including the s-wave component and backgrounds for each direction of the neutron polarization. The asymmetry is, therefore, dependent on the backgrounds. Since they used a NaI detector with poor energy resolution, they could not identify the photo-peak derived from background $\gamma$-rays, including the background-origin. The asymmetry is underestimated due to background $\gamma$-rays emitted from other nuclei.

In this paper, a more precise measurement of the transverse asymmetry of the 1.3-eV resonance in the $^{117}$Sn(n,$\gamma$)$^{118}$Sn reaction at the J-PARC is reported. The neutron source at the J-PARC has better energy resolution than that of a reactor like IBR-30, making them suitable for precision measurements. The transverse asymmetry without backgrounds was defined, and its angular dependence was measured using Ge detectors with high energy resolution and the $^3$He spin filter with low neutron backgrounds. In particular, precise evaluation of $a_2$ term can be performed by measuring the angular dependence of the transverse asymmetry.

\section{Experiment}
\subsection{Experimental setup}
The experiment was carried out at beamline 04 of the MLF in the J-PARC. The transverse asymmetry for the 1.3-eV resonance of $^{117}$Sn was measured with the same setup as the past measurement for the 0.74-eV resonance of $^{139}$La \cite{Yamamoto}. In the MLF, the pulsed proton beams accelerated by the 3-GeV Rapid-Cycling Synchrotron of the J-PARC produce the pulsed neutrons by a spallation reaction in a mercury target. The proton beams with a double-bunch structure and an average of 615 kW power were incident upon to the spallation target with a specific repetition ratio of 25 Hz. Therefore, the neutron energy can be determined from the neutron time-of-flight, TOF.

\begin{figure}[htbp]
\includegraphics[clip,width=8cm]{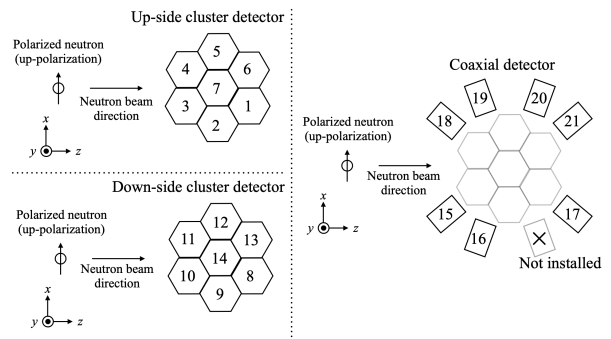}
\caption{\label{detnum}Ge detector arrangement and detector numbers. A hexagon and rectangle indicate Ge crystals, and the number in the hexagon or rectangle is the detector number.}
\end{figure}

\begin{figure}[htbp]
\includegraphics[clip,width=6cm]{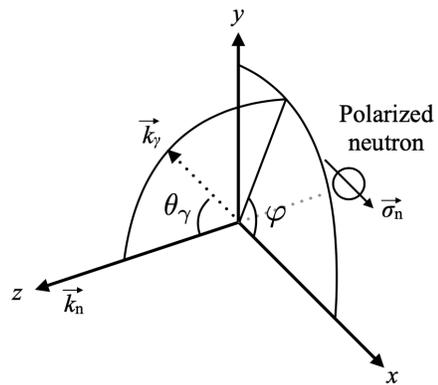}
\caption{\label{angdef}Definition of the angles and axes. $\vec{k}_\gamma$ and $\vec{k}_\textrm{n}$ represent the directions of the emitted $\gamma$-ray and incident neutron. The angle $\theta_\gamma$ is defined as the angle between the $\gamma$-ray and neutron directions in the perpendicular plane to the direction of neutron polarization ($0<\theta_\gamma<\pi$). $\varphi$ is the angle between the $x$-axis and $\gamma$-ray direction ($0<\varphi<2\pi$).}
\end{figure}

The Ge $\gamma$-ray detector assembly was installed at 21.5-m flight length. There were two types of Ge detectors, cluster and coaxial types. Seven coaxial detectors were arranged to surround the sample on the same plane as the neutron beam. The cluster detectors, bundled with seven Ge detectors, were located at the up-side and down-side of the sample. More details of the beamline are given in Ref. \cite{Igashira,Takada}. Figure \ref{detnum} depicts the detector number definitions. 
The angles, $\theta_\gamma$ and $\varphi$, were defined as shown in Fig. \ref{angdef}, and the angles of each cluster type detector are listed in Table. \ref{detangle}. The angles of all coaxial type detectors were $\varphi=0$.
\begin{table}[htbp]
\caption{\label{detangle}
Angles of each detector}
\begin{ruledtabular}
\begin{tabular}{ccc|ccc}
\multicolumn{3}{c}{up-side detector} & \multicolumn{3}{c}{down-side detector} \\
detector & $\theta_\gamma$ (deg) & $\varphi$ (deg) & detector & $\theta_\gamma$ (deg) & $\varphi$ (deg)\\
\hline
1 & 70.9 & 101.8 & 8 & 70.9 & 258.2  \\
2 & 90 & 113.7  & 9 & 90 & 246.3  \\
3 & 109.1 & 101.8  & 10 & 109.1 & 258.2 \\
4 & 109.1 & 78.2& 11 & 109.1 & 281.8 \\
5 & 90 & 66.3  & 12 & 90 & 293.7 \\
6 & 70.9 & 78.2 & 13 & 70.9 & 281.8\\
7 & 90 & 90  & 14 & 90 & 270 
\end{tabular}
\end{ruledtabular}
\end{table}

The differential cross-section in Eq. (\ref{eq:cor1}) can be written in the laboratory system as shown below:
\begin{eqnarray}
\label{eq:cor}
    \frac{d\sigma_{\textrm{n}\gamma}(\theta_\gamma,\phi)}{d\Omega}&&=\frac{1}{2}\left\{a_0+a_1\cos\theta_\gamma-a_2P_\textrm{n}\sin\theta_\gamma\sin\varphi\right.\\ \nonumber
    &&\left.+a_3\left(\cos^2\theta_\gamma-\frac{1}{3}\right)\right\}.
\end{eqnarray}
Thus, the effect of the $a_2$ term vanishes in coaxial type detectors due to $\varphi=0$. V1724 (14 bits, 100 MHz) modules and the CoMPASS software supported by Costruzionl Apparecchiature Elettroniche Nucleari SpA (CAEN) were used to acquire the difference between proton incident timing and the $\gamma$-ray detection timing, $t^{m}$, corresponding to TOF, and the pulse height corresponding to the deposited $\gamma$-ray energy, $E_{\gamma}$, in a list-mode.

The $^{3}$He neutron spin filter was used to produce polarized neutrons. The neutrons are polarized by passing through the $^{3}$He spin filter because the neutron capture cross-section of $^{3}$He strongly depends on the spin direction. The $^{3}$He nuclei were polarized using the Spin Exchange Optical Pumping (SEOP) method with rubidium and potassium \cite{Earl2003}. The details of the $^{3}$He spin filter at the J-PARC are described in Ref. \cite{Okudaira2018}.

The transmitted neutrons were monitored by two types of Li-glass detectors installed at 28.8-m flight length to determine the neutron polarization ratio. One consisted of a $^{6}$Li enriched ($\ge$95\%) Li-glass scintillator, GS20, from Saint-Gobain S.A. and PMT H7195 from Hamamatsu Photonics K.K. This detector can detect neutrons via the $^{6}$Li(n,$\alpha$)t reaction. The other was a $^{7}$Li enriched ($\ge$99.9\%) Li-glass scintillator, GS30, with slight sensitivity to neutrons because of the small neutron cross-section of $^{7}$Li. The $\gamma$-ray backgrounds were thus subtracted using neutron spectrum obtained by the $^7$Li enriched detector. A V1720 (12 bits, 250 MHz) module and the CoMPASS software were used for data acquisition.

\subsection{Measurement}
The natural Sn metal with 40 mm $\times$ 40 mm $\times$ 6 mm was used as the sample. The measurement times for the up- and down-polarized neutrons were 20.5 h and 20.8 h, respectively. The transmission for unpolarized $^{3}$He was also measured for 2.0 h to determine the polarization ratio of $^{3}$He.

\section{Analysis and Results}
\subsection{Definition of transverse asymmetry}
\begin{figure}[htbp]
\includegraphics[clip,width=8cm]{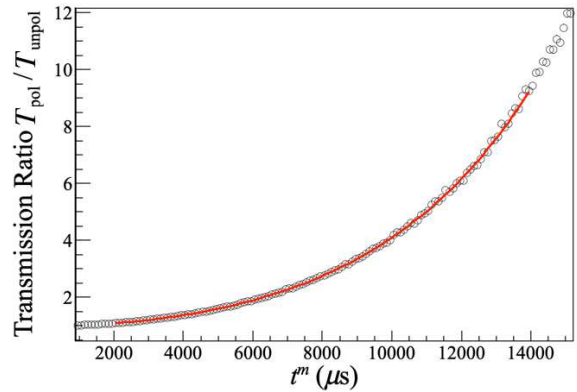}
\caption{\label{henkyokuritu}Ratio obtained by dividing the transmission through the polarized $^{3}$He cell by that through the unpolarized one.}
\end{figure}

\begin{figure}[htbp]
\includegraphics[clip,width=8cm]{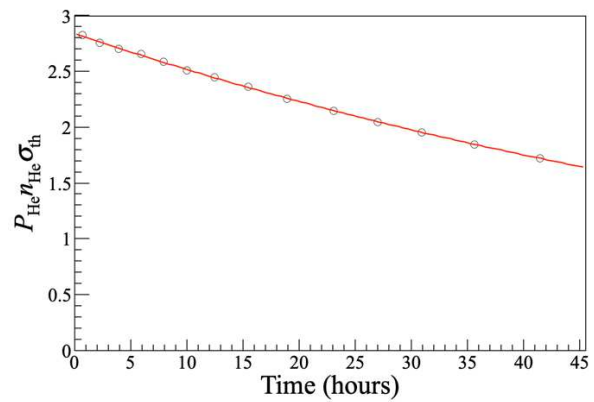}
\caption{\label{relaxtime} Time dependence of the $^{3}$He polarization ratio.}
\end{figure}

The neutron polarization ratio can be obtained from the transmission measurement. Figure \ref{henkyokuritu} shows the ratio of the neutron TOF spectrum transmitted through the polarized $^{3}$He cell divided by that through the unpolarized one. 
The polarization ratio of $^{3}$He, defined as $P_\textrm{He}$, was determined by fitting the ratio with the following function:
\begin{equation}
T_\textrm{pol}/T_\textrm{unpol}=\cosh(P_\textrm{He}n_\textrm{He}\sigma_\textrm{th}v_\textrm{th}/v),
\end{equation}
where $n_\textrm{He}$ is the areal density of $^{3}$He; $v$ is the neutron velocity; $v_\textrm{th}$ and $\sigma_\textrm{th}$ are the velocity and the capture cross-section of $^{3}$He at the thermal-neutron energy, respectively. The solid line in Fig. \ref{henkyokuritu} shows the fitting results, and the average $^{3}$He polarization ratio was determined as 75\% for the first 1.5 h. The $^{3}$He polarization relaxes as time proceeds because of the nonuniformity of external magnetic fields or collisions between $^{3}$He atoms and so on. Figure \ref{relaxtime} presents the $^{3}$He polarization ratio against the time, and the solid line represents the fitting results by the following function:
\begin{equation}
 P_\textrm{He}(t)n_\textrm{He}\sigma_\textrm{th}=n_\textrm{He}\sigma_\textrm{th}P_{\textrm{He},0}\exp(-t/\tau),
\end{equation}
where $t$ is time; and $\tau$ is the relaxation time. 
The relaxation time of the $^{3}$He was $82.65 \pm 0.07$ h. The neutron polarization ratio during the up-polarization measurement was obtained as follows:
\begin{eqnarray}
\overline{P}^\textrm{up}_\textrm{n}&=&\int_{T_\textrm{up}}P_\textrm{n}(t) dt/T_\textrm{up} \nonumber \\
&=&\int_{T_\textrm{up}}\tanh(n_\textrm{He}\sigma_\textrm{th}P_\textrm{He,0}\exp(-t/\tau)v_\textrm{th}/v)dt/T_\textrm{up}.
\end{eqnarray}
The polarization ratios during measurements for each polarization direction were $\overline{P}_\textrm{n}^\textrm{up}=0.291 \pm 0.001$ and $\overline{P}_\textrm{n}^\textrm{down}=0.296 \pm 0.001$ at the 1.3-eV resonance region. 

The asymmetry is defined as:
 \begin{equation}
 \varepsilon_{\gamma,d}=\frac{n_d^\textrm{up}-n_d^\textrm{down}}{n_d^\textrm{up}+n_d^\textrm{down}},
 \end{equation}
 where $n^\textrm{up}_d$ and $n^\textrm{down}_d$ are the numbers of events at detector $d$ in each neutron polarization direction. The transverse asymmetry of each detector $d$ is defined as:
 \begin{equation}
 A'_{\textrm{LR},d}=\frac{2\varepsilon_{\gamma,d}}{(P_\textrm{n}^\textrm{up}+P_\textrm{n}^\textrm{down})-\varepsilon_{\gamma,d}(P_\textrm{n}^\textrm{up}-P_\textrm{n}^\textrm{down})}.
 \end{equation}
 Since the capture events after scattering in the Sn sample are not dependent on the polarization direction, the corrected asymmetry $A_{\textrm{LR},d}$ can be written \cite{Yamamoto} as:
\begin{equation}
A_{\textrm{LR,d}}=A'_{\textrm{LR},d}\left(1+\frac{n_\textrm{sct}}{n_\textrm{res}}\right),
\end{equation} 
where $n_\textrm{sct}$ is the capture events after scattering, and $n_\textrm{res}=(n^\textrm{up}+n^\textrm{down})/2$ is the number of capture reaction events at the resonance.
 
\begin{figure}[htbp]
\includegraphics[clip,width=8cm]{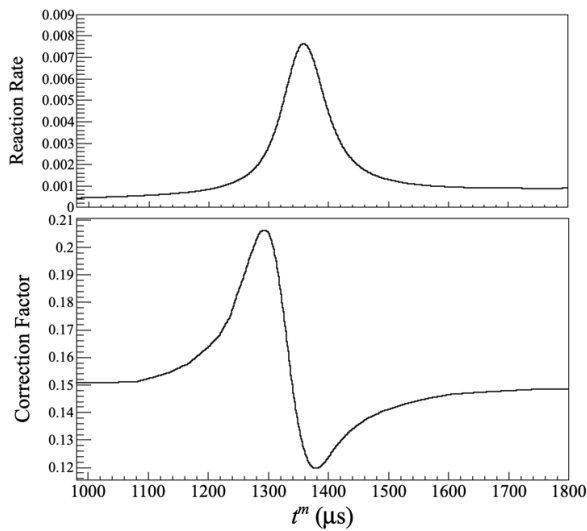}
\caption{\label{scatcorfac}Capture reaction rate $n_\textrm{res}$ (top) and correction factor $n_\textrm{sct}/n_\textrm{res}$ (bottom) obtained by PHITS simulation. The reaction rate corresponds to the capture cross-section considering effects of self-shielding and multiple scattering in the sample.}
\end{figure}

The correction factor $n_\textrm{sct}/n_\textrm{res}$ was calculated using the Monte Carlo simulation code, PHITS \cite{PHITS}. The resolution function of beamline 04 obtained by Kino et al. \cite{Kino} and Doppler broadening were considered in the correction factor, $n_\textrm{sct}/n_\textrm{res}$, as shown in Fig. \ref{scatcorfac} with the capture reaction rate, $n_\textrm{res}$, which corresponds to the capture cross-section considering effects of self-shielding and multiple scattering.
Notably, the correction factor depends on the neutron energy because neutrons with higher energies than resonance are captured by the decrease in energy due to scattering. Thus, the correction factor was increased at higher energy, or lower TOF, than the resonance.

\subsection{Transverse asymmetry at the 1.3-eV resonance of $^{117}$Sn}
\begin{figure}[htbp]
\includegraphics[clip,width=8cm]{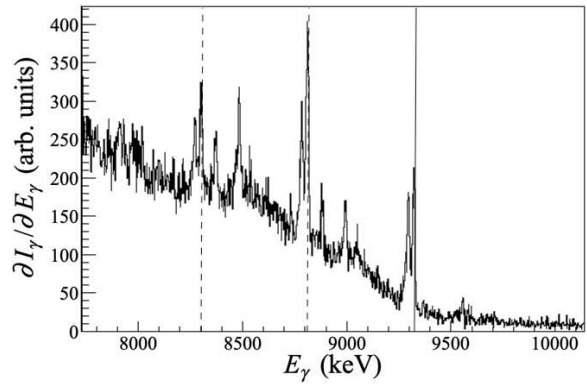}
\caption{\label{gammaspe}$\gamma$-ray deposit energy spectrum of detector 1 in Sn measurement.}
\end{figure}

\begin{figure}[htbp]
\includegraphics[clip,width=8cm]{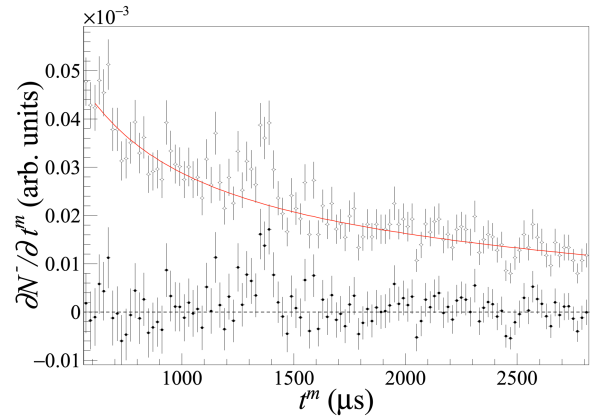}
\caption{\label{gatetof} Gated histogram of $\gamma$-ray counts as a function of $t^{m}$. The open circle represents the histogram of detector 14 in Sn measurement using down-polarized neutron, and the closed circle represents that after background subtraction. The solid line is the fitting result of the background.}
\end{figure}

Figure \ref{gammaspe} depicts the obtained $\gamma$-ray deposit energy spectrum of detector 1 in Sn measurement. 
The vertical solid line indicates the 9327.5 keV transition to the ground-state of $^{118}$Sn from the compound state of $^{117}$Sn+n, and the dotted lines represent its single and double escape peaks. The neutron capture reaction of $^{115}$Sn is responsible for the 9563.1 keV peak. The other peaks are derived from the scattered neutron capture reaction of iron contained in the guide magnet and shield of the Ge detector assembly. Since the transition to the first excited state of $^{118}$Sn from the compound-state emits 8069.8 keV $\gamma$-ray, it is considered that $\gamma$-rays above the double escape peak, 8305.5 keV photo-peak, contain the only transition to the ground-state of $^{118}$Sn, including backgrounds derived from other nuclei. Furthermore, backgrounds by iron and $^{115}$Sn do not affect the transverse asymmetry because their cross-sections have no resonance or unique structure around 1.3 eV. Therefore, the $\gamma$-ray gate region was set from 8250 keV to 9400 keV, and Fig. \ref{gatetof} shows the gated histogram of $\gamma$-ray counts as a function of $t^{m}$ with the open circle.
To remove the s-wave component of $^{117}$Sn and the backgrounds from other nuclei, the histogram was fitted using the following function:
\begin{equation}
    f(t^{m})=c_0+c_1t^{m}+c_2/t^{m},
\end{equation}
with $700\ \mu s\le t^{m} \le 1200\ \mu s$ and $1700\ \mu s\le t^{m} \le 2500\ \mu s$. Figure \ref{gatetof} presents the fitting result with the solid line. The histogram after subtracting the fitting function is also shown in Fig. \ref{gatetof} with the closed circle.

\begin{figure}[htbp]
\begin{minipage}{7cm}
\includegraphics[clip,width=7cm]{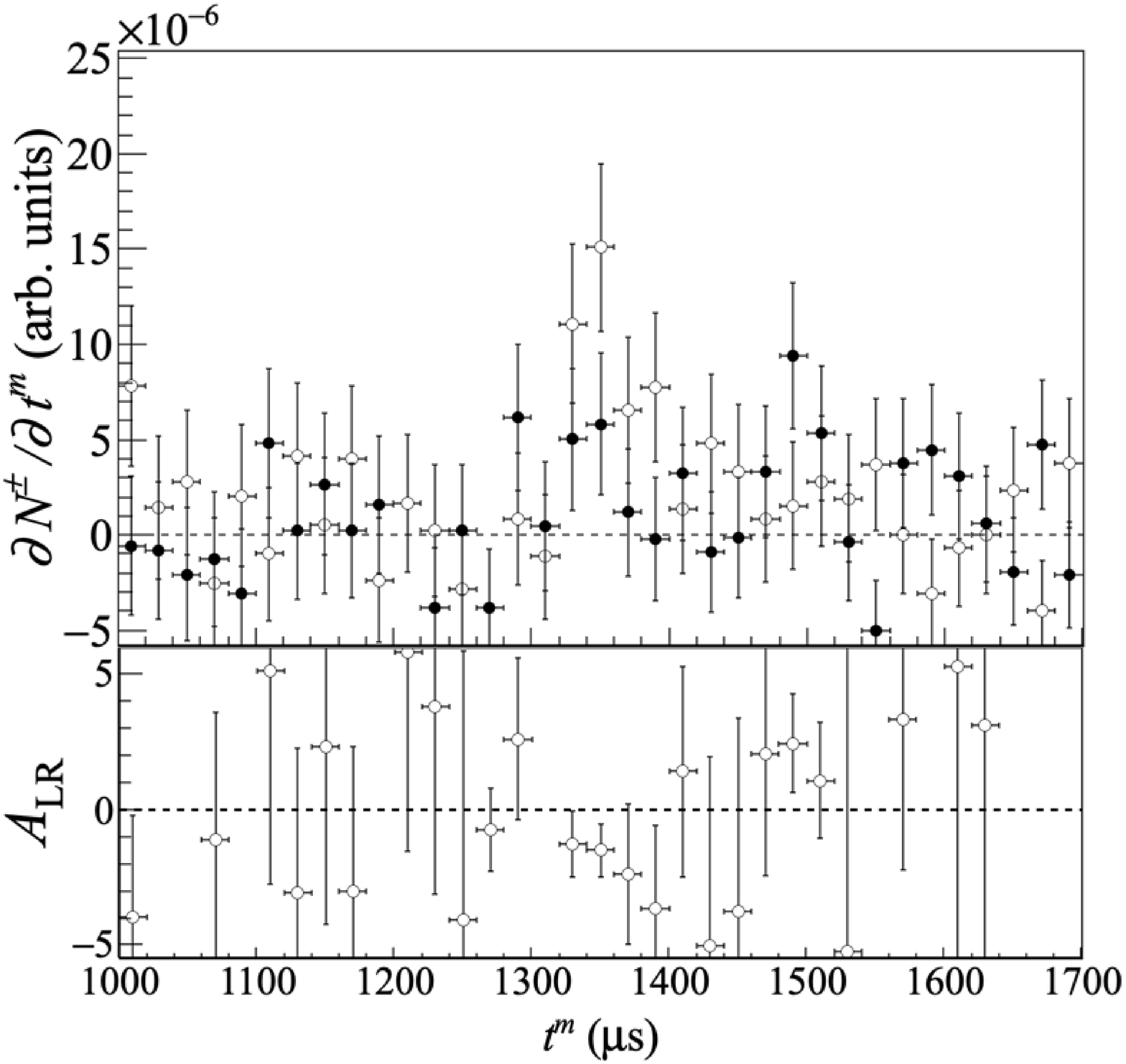}
(a) detector 1
\end{minipage}
\\
\begin{minipage}{7cm}
\includegraphics[clip,width=7cm]{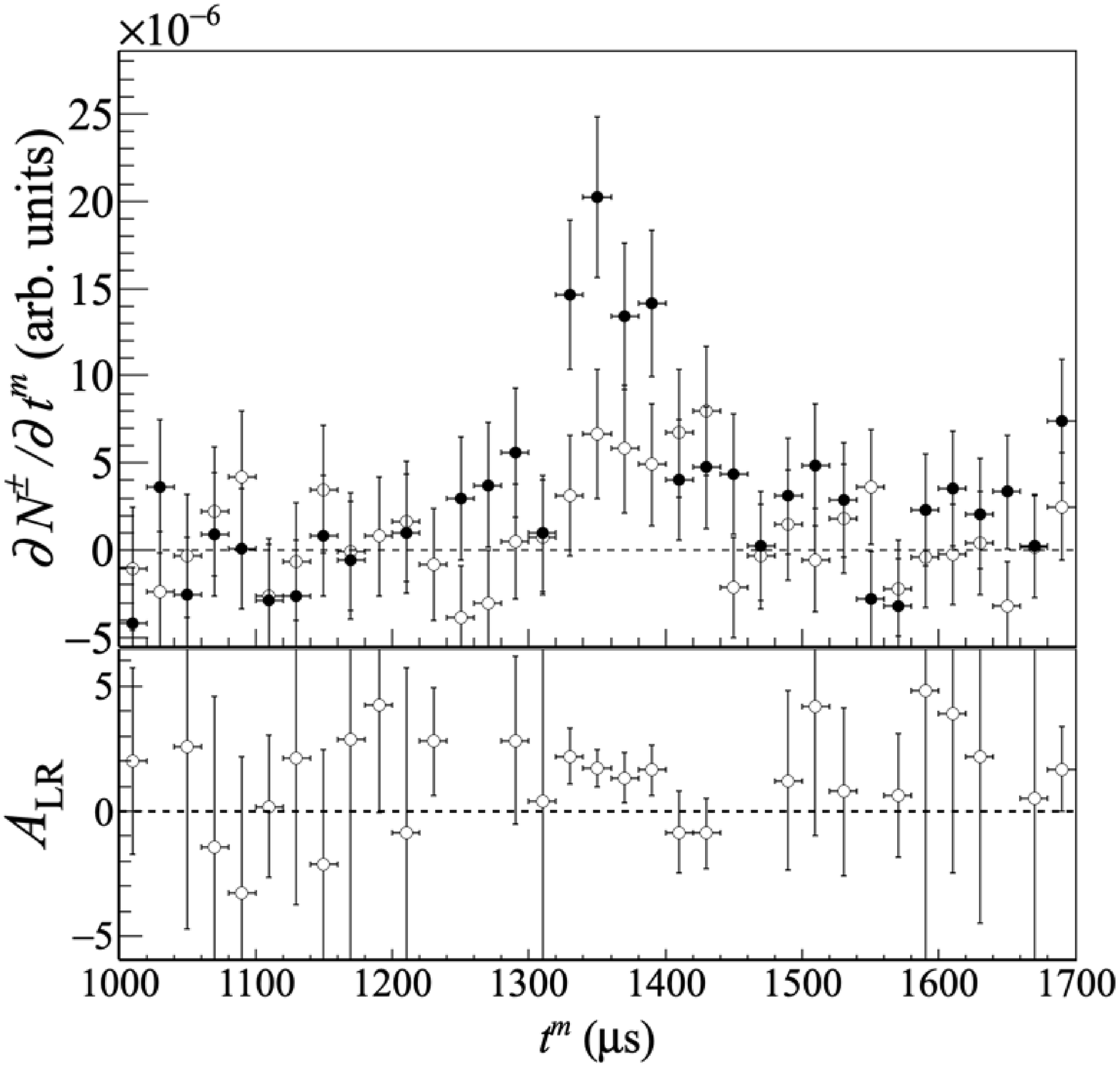}
(b) detector 14
\end{minipage}
\caption{\label{updownasym}$\gamma$-ray counts for each polarization direction after background subtraction and $A_\textrm{LR}$. The closed and open circles indicate for the up- and down-polarization measurements in the upper side figure. The sign of $A_\textrm{LR}$ is reversed for the up-side detector (detector 1) and down-side detector (detector 14).}
\end{figure}

Figure \ref{updownasym} shows the histograms for each polarization direction after background subtraction and calculated transverse asymmetry $A_{\textrm{LR}}$.
Table \ref{ALReach} lists $A_\textrm{LR}$ in a neutron energy region $E_\textrm{p}-2\Gamma_\textrm{p}\leq E_\textrm{n}\leq E_\textrm{p}+2\Gamma_\textrm{p}$ for each detector. 
\begin{table*}[htbp]
\caption{\label{ALReach}
$A_\textrm{LR}$ for each detector in a neutron energy region $E_\textrm{p}-2\Gamma_\textrm{p}\leq E_\textrm{n}\leq E_\textrm{p}+2\Gamma_\textrm{p}$. Detectors 7 and 16 were not used because the $\gamma$-ray energy resolution was poor due to electrical noises. $A^{90^\circ}_{\textrm{LR},d}=A_{\textrm{LR},d}/(\sin\theta_\gamma\sin\varphi)$ is the asymmetry converted to $90^\circ$.} 
\begin{ruledtabular}
\begin{tabular}{ccc|ccc|cc}
\multicolumn{3}{c}{up-side detector} & \multicolumn{3}{c}{down-side detector} & \multicolumn{2}{c}{coaxial detector}\\
detector & $A_{\textrm{LR},d}$ & $A^{90^\circ}_{\textrm{LR},d}$ & detector & $A_{\textrm{LR},d}$ & $A^{90^\circ}_{\textrm{LR},d}$ & detector & $A_{\textrm{LR},d}$\\
\hline
1 & $-0.46\pm0.96$ & $-0.50\pm1.04$ & 8 & $0.62\pm0.50$ & $-0.67\pm0.54$ & 15 & $-0.04\pm3.18$ \\
2 & $-1.73\pm0.92$ & $-1.89\pm1.00$ & 9 & $1.06\pm0.74$ & $-1.16\pm0.81$ & 16 & Not used\\
3 & $-2.93\pm1.60$ & $-3.16\pm1.73$ & 10 & $1.51\pm0.77$ & $-1.63\pm0.83$ & 17 & $0.49\pm0.68$\\
4 & $-0.05\pm0.67$ & $-0.05\pm0.73$ & 11 & $1.40\pm0.78$ & $-1.52\pm0.84$ & 18 & $-0.04\pm0.80$\\
5 & $-0.18\pm0.93$ & $-0.20\pm1.02$ & 12 & $-0.19\pm0.86$ & $0.20\pm0.94$ & 19 & $1.59\pm1.27$\\
6 & $-1.02\pm0.71$ & $-1.11\pm0.76$ & 13 & $2.18\pm0.67$ & $-2.35\pm0.73$ & 20 & $0.57\pm0.68$\\
7 & Not used & - & 14 & $1.44\pm0.85$ & $-1.44\pm0.85$ & 21 & $-0.58\pm0.65$\\
Average & -  & $-0.82\pm0.38$ & Average &  - & $-1.21\pm0.29$ & Average & $0.21\pm0.33$\\
\end{tabular}
\end{ruledtabular}
\end{table*}
Here, the resonance parameters were used, as shown in Table \ref{respara}.
\begin{table}[htbp]
\caption{\label{respara}
Resonance parameters of $^{117}$Sn}
\begin{ruledtabular}
\begin{tabular}{cccc}
$E_\textrm{r}$ [eV] & $J/l$ & $\Gamma_\gamma$ [meV] & $g\Gamma_\textrm{n}$ [meV] \\ \hline
$1.331\pm0.002\footnote[1]{taken from Ref. \cite{Koga}}$ & 1/1\footnote[2]{taken from Ref. \cite{Smith}} & $133\pm5$\footnotemark[1] & $1.38\times10^{-4}$\footnotemark[2]  \\
\end{tabular}
\end{ruledtabular}
\end{table}
The $A^{90^\circ}_{\textrm{LR},d}$ is the transverse asymmetry converted to 90$^\circ$, calculated by $A^{90^\circ}_{\textrm{LR},d}=A_{\textrm{LR},d}/(\sin\theta_\gamma\sin\varphi)$. It is found that the average of $A^{90^\circ}_{\textrm{LR},d}$ is consistent between up- and down-side detectors. The transverse asymmetry averaged over up- and down-side detectors was obtained as $A^{90^\circ}_{\textrm{LR}}=-1.07 \pm 0.23$. Furthermore, the average of $A_{\textrm{LR},d}$ of the coaxial detector was consistent with 0 due to $\sin\varphi = 0$.

\section{Discussion}
\subsection{Comparison to the previous study}
\begin{figure}[htbp]
\includegraphics[clip,width=8cm]{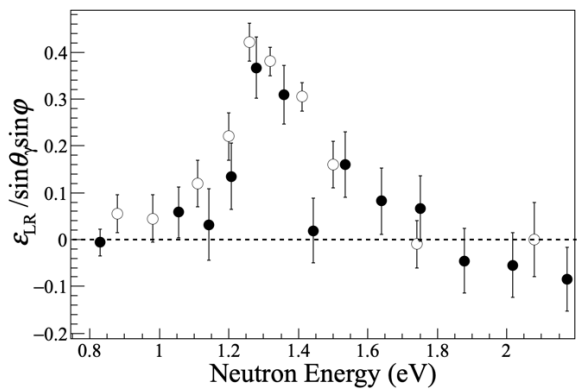}
\caption{\label{epsilon}Left-right asymmetry defined same as in the previous study. The closed and open circles represent the present and previous results \cite{Skoi}, respectively.}
\end{figure}

In the previous study of the left-right asymmetry measurements of $^{117}$Sn, the asymmetry $\varepsilon_\textrm{LR}$ was reported in Fig. 13 in Ref. \cite{Skoi}. Notably, the asymmetry $\varepsilon_\textrm{LR}$ was defined, including the s-wave components around the p-wave resonance. Conversely, the aforementioned analysis removed the s-wave components with backgrounds in the present definition of $A_\textrm{LR}$. To compare the present result with the previous study, $\varepsilon_\textrm{LR}$ was also calculated and plotted in Fig. \ref{epsilon} with the previous results. In the present analysis, the $\varepsilon_\textrm{LR}$ was calculated for all detectors and divided by $\sin\theta_\gamma\sin\varphi$, and average values of all detectors were plotted in Fig. \ref{epsilon}, consistent with the previous study, which implies that the background conditions for our and Skoy's cases were comparable. The result of $\gamma$-ray identification suggested that the asymmetry was influenced by backgrounds such as iron isotopes. Therefore, it was necessary to evaluate the asymmetry without backgrounds, $A_\textrm{LR}$.

\subsection{Angular dependence of the transverse asymmetry}
\begin{figure}[htbp]
\includegraphics[clip,width=8cm]{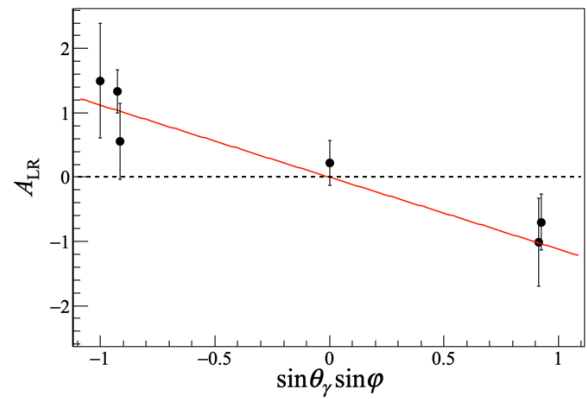}
\caption{\label{angle}Angular dependence of the transverse asymmetry. The solid line represents the fitting results. Each point plots the weighted average of $A_\textrm{LR}$ for detector 14 for $\sin\theta_\gamma\sin\varphi=-1$, detectors 8, 10, 11, and 13 for $-0.925$, detectors 9 and 12 for $-0.916$, detectors 15 through 21 for $0$, detectors 2 and 5 for $0.916$, and detectors 1, 3, 4, and 6 for $0.925$, respectively.}
\end{figure}

To discuss the angular dependence of $a_2$ term, $A_\textrm{LR}$ is plotted for $\sin\theta_\gamma\sin\varphi$ in Fig. \ref{angle}. The weighted average of $A_\textrm{LR}$ was calculated and plotted for each angle. According to Eq. (\ref{eq:cor}), the transverse asymmetry could be written as:
\begin{equation}
    A_\textrm{LR}=\frac{-A_{2}\sin\theta_\gamma\sin\varphi}{1+A_3(\cos^2\theta_\gamma-1/3)},
\end{equation}
where $A_i$ is
\begin{equation}
    A_i=\int_{E_\textrm{p}-2\Gamma_\textrm{p}}^{E_\textrm{p}+2\Gamma_\textrm{p}}a_idE_\textrm{n}/\int_{E_\textrm{p}-2\Gamma_\textrm{p}}^{E_\textrm{p}+2\Gamma_\textrm{p}}a_{0,p}dE_\textrm{n}.
\end{equation}
Here, $a_{0,\textrm{p}}$ is the $a_0$ term only considering the 1.4-eV p-wave resonance. 
Furthermore, $a_1$ term can be ignored by integral in Eq. (11) because its energy-dependence is odd function of energy centered at the p-wave resonance \cite{Okudaira,Koga}.
In this study, the angular dependence of the transverse asymmetry was little sensitive to $A_3$ because the Ge detectors were placed around $\cos^2\theta_\gamma\sim0$. If the angular dependence of the $a_3$ term was ignored, the angular dependence of the transverse asymmetry can be written as:
\begin{equation}
\label{angulardependence}
    A_\textrm{LR}=-\tilde{A}_\textrm{LR}\sin\theta_\gamma\sin\varphi,
\end{equation}
where $\tilde{A}_\textrm{LR}=\frac{A_2}{1-A_3/3}$.
The solid line in Fig. \ref{angle} shows the fitting results by Eq. (\ref{angulardependence}), and $\tilde{A}_\textrm{LR}=1.07\pm0.23$ was obtained. The transverse asymmetry depending on $\sin\theta_\gamma\sin\varphi$ was evaluated, and it is caused by the $a_2$ term.

\section{Conclusion}
In this study, the transverse asymmetry, independent of backgrounds, for $^{117}$Sn(n,$\gamma$)$^{118}$Sn was measured using Ge detectors installed at several angles in the J-PARC beamline 04, and a non-zero asymmetry was obtained. 
The transverse asymmetry corresponding to $a_2$ term was also evaluated on its angular dependence.
In the future, the validation of the s-p mixing model will be performed by global analysis using present result and other terms.


\begin{acknowledgments}
The authors would like to thank the staff of beamline 04 for the maintenance of the germanium detectors, and the MLF and J-PARC for operating the accelerators and the neutron production target. The neutron experiments at the MLF of J-PARC were performed under the user program (Proposals No. 2019A0185). This work was supported by the Neutron Science Division of KEK as an S-type research project with program number 2018S12. This work was partially supported by MEXT KAKENHI Grant No. JP19GS0210 and JSPS KAKENHI Grant No. JP17H02889.
\end{acknowledgments}


\end{document}